\documentclass[12pt]{article}

\usepackage{a4wide,graphics,graphicx,amsmath,amssymb,cite,nicefrac}
\usepackage[verbose]{wrapfig}
\usepackage{authblk,hyperref}
\catcode`@=11 \@addtoreset{equation}{section} \catcode`@=12
\begin{document}
\date{}
\title{{New supersymmetric black holes in four dimensional $N=2$ supergravity}
{\baselineskip -.2in
} 
\vbox{
{\bf \LARGE }
}}
\author{Taniya Mandal\thanks{email: taniya@physics.iitm.ac.in} } 
\author{Prasanta K. Tripathy\thanks{email: prasanta@iitm.ac.in}}
\affil{\normalsize\it Department of Physics, \authorcr \it Indian Institute of Technology Madras, \authorcr \it Chennai 600036, India.}
\maketitle
\begin{abstract}
In this paper we consider the four dimensional $N=2$ supergravity theory arising from the compactification of type $IIA$ string  theory
on a Calabi-Yau manifold. We analyse the supersymmetric flow equations for static, spherically symmetric, single-centered black 
holes. These flow equations are solved by a set of algebraic equations involving the holomorphic sections and harmonic functions. 
We examine black hole configurations with $D0-D4-D6$  charge for which the most general solution of these algebraic equations 
are considered. Though the black hole solution is unique for a given value of the charges, we find new phases of the black hole 
solutions upon varying them.
\end{abstract}

\newpage

The study of black holes has remained to be an interesting area of research for the past several decades because of the long 
standing problems, such as the origin of black hole entropy and the information loss paradox associated with them.  The issue
pertaining to the black hole entropy is well understood in the context of supersymmetry preserving black hole solutions arising
in string theory \cite{Strominger:1996sh}.  In the microscopic analysis the states are described in terms of intersecting $D$-branes. 
Entropy associated with the supersymmetric multiplets are determined in terms of the central charge of the corresponding world 
volume theory by using the Cardy formula \cite{Cardy:1986ie}. These intersecting D-branes can be wrapped on appropriate 
cycles of a Calabi-Yau manifold to give rise four dimensional black hole solutions in the large volume limit. Entropy of these 
black holes are computed by the so called attractor mechanism\cite{Ferrara:1995ih,Strominger:1996kf}. These entropies match 
with the corresponding ones obtained by using the microscopic study of the world volume theory.

The attractor mechanism has another important feature apart from its role in determining the entropy of the macroscopic black
hole solutions, which is related to the uniqueness of the solution for a given value of charges \cite{Moore:1998pn,Moore:1998zu}. 
In the case of supersymmetric  attractors, for simple charge configurations it is straightforward to prove the uniqueness using the 
attractor equations explicitly\cite{Manda:2015zoa}.  In general using the extremization of the central charge one can systematically 
prove the uniqueness of the supersymmetric  attractors for a given set of charges by considering the extermization of the central 
charge along with the positivity of the moduli space metric \cite{Wijnholt:1999vk}. 

However, this should not give an impression that there is a unique attractor configuration for a given type of dyonic black hole.
There are indeed different phases of black hole attractors for a given dyonic charge configuration. This exposition can seemly 
be illustrated in the context of $N=2$ supergravity coupled to a number of vector multiplets.  
The entropy for the extremal black  holes in such theories are determined in terms of a symplectic invariant \cite{Bellucci:2006xz}. 
For a given dyonic 
configuration, the symplectic invariant is a real valued function of the black hole charges. The symplectic invariant may change 
sign as one varies the charges. The value of black hole entropy depends only on the modulus of the symplectic invariant, 
however the preservation of supersymmetry of the extremal configuration depends upon its sign \cite{Tripathy:2005qp}.
In addition to the sign of this symplectic invariant, the existence of the supersymmetric attractors depend on the positive 
definiteness of the moduli space metric and the gauge kinetic terms.

The positive definiteness of the moduli space metric and gauge kinetic terms at the attractor point imposes more conditions
on the black hole charges. Such conditions, for a given type of dyonic configuration in general divide the charge lattice into 
several domains and give rise to the possibility of the existence of new attractor points corresponding to these different 
domains. Such a possibility has been explored in a recent work by the present authors by studying the most general solution
for the attractor equations in the four dimensional supersymmetric black holes in type $IIA$ string theory compactified on a 
Calabi-Yau manifold  \cite{Manda:2015zoa}. The most general solution for the $D0-D4-D6$ attractors has been obtained 
and it has been shown that there are indeed different domains in the charge lattice and the form of the solution depends 
upon the particular domain to which the black hole charges belong. 

The goal of the present work is to establish the exact analytic expression of the full interpolating solution corresponding to 
these new phases of black hole attractors. A general technique for such constructions for axion-free black holes 
 \cite{Sabra:1997dh,  Sabra:1997kq} as well as axionic black holes \cite{Mohaupt:2000mj} has already 
 been developed. Using these results we will derive the full flow for the new phases of  supersymmetric black holes.

Our focus in this paper is entirely based on the four dimensional $N=2$ supergravity theory coupled to $n$ vector multiplets \cite{Ferrara:1997tw}
\begin{eqnarray}\label{sugra}
{\cal L} = - \frac{R}{2} +  g_{a\bar b} \partial_\mu x^a \partial_\nu \bar x^{\bar b} h^{\mu\nu} 
-  \mu_{\Lambda\Sigma} {\cal F}^\Lambda_{\mu\nu}{\cal F}^\Sigma_{\lambda\rho} 
h^{\mu\lambda}h^{\nu\rho} -   \nu_{\Lambda\Sigma} {\cal F}^\Lambda_{\mu\nu}
*{\cal F}^\Sigma_{\lambda\rho} h^{\mu\lambda}h^{\nu\rho} \ .
\end{eqnarray}
Here $R$ is the Ricci scalar corresponding to the space-time metric $h_{\mu\nu}$, $g_{a\bar b}$ is the moduli space metric,
$\mu_{\Lambda\Sigma}$ and $\nu_{\Lambda\Sigma}$ are the gauge couplings. For $N=2$ supergravity theory, the moduli 
space metric and gauge couplings are determined in terms of a pre-potential $F$.

We are interested in asymptotically flat, static, spherically symmetric solutions. The metric for such configuration is determined
in terms of a single warp factor $U(\tau)$ and has the form
\begin{equation}
 ds^2=e^{2U(\tau)}dt^2-e^{-2U(\tau)}(d\vec{x})^2 \ .
\end{equation}
The warp factor $U$ solely depends on the radial coordinate $r$. For convenience, we introduce the coordinate $\tau=1/r$.
Using the equations for the gauge fields and the above ansatz for the metric one can show that \cite{Ferrara:1997tw}, such 
a system can indeed be described in terms of an effective one dimensional theory whose Lagrangian density is given by
\begin{equation}\label{leff}
 \mathcal{L}(U, x^a(\tau),\bar{x}^a(\tau))=\bigg(\frac{dU}{d\tau}\bigg)^2+g_{a\bar b} \frac{dx^a}{d\tau} \frac{d\bar x^b}{d\tau}
 +e^{2U}(|Z|^2+|D_aZ|^2) \ .
\end{equation}
Here $Z$ denotes the central charge of the $N=2$ theory: 
\begin{equation}
Z = e^{K/2} (X^\Lambda q_\Lambda - F_\Lambda p^\Lambda) \ ,
\end{equation}
with $(q_\Lambda,p^\Lambda)$ being the dyonic charges of the black hole, $X^\Lambda$ are the symplectic sections, $F$ is
the $N=2$ pre-potential and $F_\Lambda = \partial_\Lambda F$. The symplectic sections $X^\Lambda$ are  related to the 
complex scalars $x^a$ by $x^a = X^a/X^0$. In this paper we choose the gauge $X^0=1$.  Here, $D_a$ denotes the K\"ahler 
covariant derivative $D_a Z = \partial_a Z + (1/2) \partial_a K Z$.  The K\"ahler potential $K$ is given by 
\begin{equation} \label{kahlerpot}
K =  -\log\Big[i\sum_{\Lambda=0}^{n} (\overline{X^\Lambda} \partial_\Lambda F 
- X^\Lambda \overline{\partial_\Lambda F})\Big] \ . 
\end{equation}

In this paper, we will focus on the four dimensional $N=2$ supergravity arising from type $IIA$ string theory compactified on a 
Calabi-Yau manifold. In the large volume limit, the pre-potential is given by 
\begin{equation}\label{prepot}
F=D_{abc}\frac{X^aX^bX^c}{X^0} \ . 
\end{equation}

We will now review the attractor solutions for the system. The attractor points are obtained upon solving the equation $D_aZ = 0 $.
At the horizon this gives rise to a set of algebraic equations. These algebraic equations can be recast in a nice covariant form
as \cite{Ferrara:2006yb}:
\begin{eqnarray}
 p^\Lambda+i \frac{\partial{\sqrt{ I_4(p,q)}}}{\partial q_\Lambda} &=& 2ie^{K/2}\bar{Z}X^\Lambda\,  \cr
 q_\Lambda-i \frac{\partial{\sqrt{ I_4(p,q)}}}{\partial p^\Lambda} &=& 2ie^{K/2}\bar{Z}F_\Lambda \label{eq:st}
\end{eqnarray}
Here $I_4(p,q)$ denotes the symplectic invariant quantity 
 \begin{equation}
  I_4(p,q)=-(p^0q_0+p^aq_a)^2+4\Big(q_0I_3(p)-p^0I_3(q)+\{I_3(q),I_3(p)\}\Big),
 \end{equation}
with  $I_3(p)=D_{abc}p^ap^bp^c$ and $I_3(q)=D^{abc}q_aq_bq_c$, where as  
$\{I_3(q),I_3(p)\}$ is the Poisson bracket 
$$\{I_3(q),I_3(p)\}=\frac{\partial I_3(q)}{\partial q_a}\frac{\partial I_3(p)}{\partial p^a} \ . $$
Here, $q_0,q_a, p^a$ and $p^0$ are respectively the $D0, D2,D4$ and $D6$ charges.
The stabilization equations \eqref{eq:st} can also be rewritten as \cite{Ferrara:1996dd,Behrndt:1996jn,Kallosh:2005ax},
\begin{equation}
 \begin{pmatrix} p^\Lambda \\ q_\Lambda \end{pmatrix} = i \begin{pmatrix} \bar{Z}f^\Lambda -Z \bar{f}^\Lambda \\ \bar{Z}h_\Lambda -Z \bar{h}_\Lambda \end{pmatrix} \ ,  \label{eq:nst}
\end{equation}
where $f^\Lambda$ and $h_\Lambda$ are given by $f^\Lambda=e^{K/2}X^\Lambda$ and $h_\Lambda=e^{K/2}F_\Lambda$.

In the present work we are interested in evaluating the interpolating solution for black holes carrying $D0-D4-D6$ charges. 
The most general attractor solution for such configurations has been obtained in \cite{Manda:2015zoa}. The solution is given 
in terms of an involutory matrix ${I^a}_b$ satisfying the relation $D_{abc}{I^b}_e{I^c}_f = D_{aef} $. The attractor 
points are given by
$x^a_{\rm attr} = x^a_1 + i x^a_2$, where 
\begin{eqnarray}
x_1^a &=& \frac{1}{p^0}\bigg(p^a - \frac{D-\frac{1}{2}q_0{p^0}^2}{D_c{I^c}_dp^d}{I^a}_bp^b\bigg)\ ,\label{eq:g2} \\
x_2^a &=& \frac{1}{p^0}\bigg(\ 1-{\bigg(\frac{D-\frac{1}{2}q_0{p^0}^2}{D_c{I^c}_dp^d}\bigg)}^2\ \bigg)^{1/2}{I^a}_bp^b\ . \label{eq:g1}
\end{eqnarray}
Here we use the notation $D_{ab} = D_{abc}p^c, D_a = D_{ab} p^b$ and $D = D_a p^a$ for convenience. In the following
we will obtain the full flow corresponding to these attractor points.

We will now study the flow equations and their solution. To obtain the flow equations, note that the effective one-dimensional 
Lagrangian eq.(\ref{leff}) can be rewritten as \cite{Ferrara:1997tw,Denef:2000nb}:
\begin{equation}
 \mathcal{L}=\bigg(\frac{dU}{d\tau} \pm e^U|Z|\bigg)^2+\bigg|\frac{dx^a}{d\tau} \pm 2e^U g^{a\bar{b}}\bar{\partial}_{\bar{b}}|Z|\bigg|^2\mp2\frac{d}{d\tau}\big(e^U|Z|\big)
\end{equation}
which gives rise to the following  set of first order BPS equations \cite{Denef:2000nb,Fre:1997jk,Andrianopoli:2006ub},
\begin{eqnarray}
 \frac{dU}{dr} &=& \pm \frac{e^U}{r^2}|Z|\label{eq:fo1}\\
\frac{dx^a}{dr} &=&  \pm 2\frac{e^U}{r^2}g^{a\bar{b}}\bar{\partial}_{\bar{b}}|Z|\label{eq:fo2}
\end{eqnarray}
The above  first order equations can also be recast as \cite{Sabra:1997dh},
\begin{eqnarray}
 \frac{dU}{dr} &=& \pm \frac{e^U}{r^2}|q_\Lambda f^\Lambda -h_\Lambda p^\Lambda|\\
\frac{dx^a}{dr} &=&  \mp \frac{e^U}{r^2} \frac{Z}{|Z|}g^{a\bar{b}}\bar{f}^\Lambda_{\bar{b}}(\mathcal{N}-\overline{\mathcal{N}})_{\Lambda\Sigma}t^\Sigma\label{eq:nfo}
\end{eqnarray}
where $t^\Lambda(r) = 
\frac{1}{2}\Big(p^\Lambda+i {(Im\mathcal{N})^{-1}}^{\Lambda\Sigma}(q_\Sigma-(Re\mathcal{N})_{\Sigma\Gamma}p^\Gamma)\Big)$ 
and $f^\Lambda_a=(\partial_a+\frac{1}{2}\partial_a K)f^\Lambda$.

The above set of first order coupled differential equations can be solved by a set of algebraic equations which have very similar 
form as that of the attractor equations. For the axion free attractors this has been carried out in 
\cite{Sabra:1997dh,Sabra:1997kq}. The axionic solution has been suggested in \cite{Mohaupt:2000mj} 
using group theoretic arguments. To express these algebraic equations, we introduce the harmonics $H^\Lambda$ and 
$H_\Lambda$ such that $H^\Lambda = \tilde h^\Lambda + \nicefrac{p^\Lambda}{r}$ and 
$H_\Lambda = \tilde h_\Lambda + \nicefrac{q_\Lambda}{r}$. Here $\tilde h^\Lambda$ and $\tilde h_\Lambda$ are arbitrary
constants satisfying $q_\Lambda\tilde{h}^\Lambda=p^\Lambda\tilde{h}_\Lambda$.  In addition, we define $Z(H)=e^{K/2}(H_\Lambda X^\Lambda-H^\Lambda F_\Lambda)$.
Further, we introduce $I_4(H) = I_4(H^\Lambda,H_\Lambda)$ by simply replacing $(p^\Lambda,q_\Lambda)$ by 
$(H^\Lambda,H_\Lambda)$ in the expression for $I_4(p,q)$. Now consider the ansatz\cite{Mohaupt:2000mj},
\begin{equation}
 e^{-2U}= |Z(H)|^2\label{eq:mans}
\end{equation}
With the help of the above ansatz it is in fact possible to show that the BPS flow equations can be solved by the following 
stabilisation equations 
 \begin{equation}
 \begin{pmatrix} H^\Lambda \\ H_\Lambda \end{pmatrix} = i \begin{pmatrix} \bar{Z}(H)f^\Lambda -Z(H) \bar{f}^\Lambda 
 \\ \bar{Z}(H)h_\Lambda -Z(H) \bar{h}_\Lambda \end{pmatrix}\label{eq:ngst}
\end{equation}

Using the general procedure developed in  \cite{Sabra:1997dh,Sabra:1997kq} we will now show that the 
above algebraic relation indeed solve the equations of motion. In the following, we will further rewrite the above equations in a 
form which will be useful for our purpose. 
From the relation $q_\Lambda\tilde{h}^\Lambda - p^\Lambda\tilde{h}_\Lambda = 0 $ we find, 
$q_\Lambda H^\Lambda-p^\Lambda H_\Lambda=0$. Using this relation and eq.(\ref{eq:ngst}) we can show that the quantity 
$Z\bar{Z}(H)$ is real. Here $Z = Z(p,q)$ is the central charge. Using eq.\eqref{eq:ngst}, and the special geometry relation 
$f^\Lambda \frac{dh_\Lambda}{dr}-\frac{df^\Lambda}{dr}h_\Lambda=0$ it can be shown that
\begin{eqnarray}
  H_\Lambda \frac{df^\Lambda}{dr}-H^\Lambda \frac{d h_\Lambda}{dr} &=& i Z(H) \Big(\bar{f}^\Lambda\frac{d h_\Lambda}{dr}-\frac{df^\Lambda}{dr}\bar{h}_\Lambda\Big)\,\cr
  H_\Lambda \frac{d\bar{f}^\Lambda}{dr}-H^\Lambda \frac{d \bar{h}_\Lambda}{dr} &=& i \bar{Z}(H) \Big(\frac{d\bar{f}^\Lambda}{dr} h_\Lambda-f^\Lambda\frac{d\bar{h}_\Lambda}{dr}\Big)\, \label{eq:cond1}
\end{eqnarray}
Now consider metric ansatz  \eqref{eq:mans}. Differentiating with respect to $r$ on both sides we find 
\begin{equation}
 2e^{-2U}\frac{dU}{dr}=\frac{1}{r^2}(\bar{Z}(H)Z+Z(H)\bar{Z})-\bar{Z}(H)\Big(H_\Lambda \frac{df^\Lambda}{dr}-H^\Lambda \frac{d h_\Lambda}{dr}\Big)-
 Z(H)\Big(H_\Lambda \frac{d\bar{f}^\Lambda}{dr}-H^\Lambda \frac{d \bar{h}_\Lambda}{dr}\Big)
\end{equation}
Using the definition of the K\"ahler potential, we find $i(\bar{f}^\Lambda h_\Lambda-f^\Lambda \bar{h}_\Lambda) = 1$. Differentiation
with respect to $r$ yields
$$\frac{d\bar{f}^\Lambda}{dr} h_\Lambda-f^\Lambda\frac{d\bar{h}_\Lambda}{dr}+\bar{f}^\Lambda\frac{d h_\Lambda}{dr}-\frac{df^\Lambda}{dr}\bar{h}_\Lambda=0 \ . $$
Using the above, along with \eqref{eq:cond1} we get $\frac{dU}{dr}=\frac{1}{r^2} e^{2U}Z(H)\bar{Z}$. Putting the metric ansatz 
\eqref{eq:mans} once more and the using reality condition of $Z(H)\bar{Z}$ in this equation we find $\frac{dU}{dr}=\frac{e^U}{r^2}|Z|$.

We would now like to demonstrate that the stabilisation equations \eqref{eq:ngst} satisfy eq.(\ref{eq:nfo}). To show this, 
note that the partial derivative of the K\"{a}hler potential (\ref{kahlerpot}) is given by 
\begin{equation}
 \partial_a K = -i e^K (\bar{X}^\Lambda \partial_a F_\Lambda - \partial_a X^\Lambda \bar{F}_\Lambda) \label{eq:pd1}
\end{equation}
Using \eqref{eq:ngst} and the covariant holomorphicity of $Z(H)$ we can express the above as 
\begin{equation}
 \partial_a K=\frac{1}{\bar{Z}(H)}(\bar{h}_\Lambda \partial_a H^\Lambda-\bar{f}^\Lambda \partial_a H_\Lambda)\label{eq:pd2}
\end{equation}
Using the above we can show that 
\begin{eqnarray}
 f^\Lambda_a \frac{dx^a}{dr}\bar{Z}(H) &=& e^{K/2}\bar{Z}(H)\frac{dX^\Lambda}{dr}+f^\Lambda (\bar{h}_\Sigma \partial_r H^\Sigma-\bar{f}^\Sigma \partial_r H_\Sigma)\, \label{eq:impeq2}
\end{eqnarray}
Consider the  imaginary part of the above equation: 
\begin{eqnarray}
 2 i {\rm Im}\Big( f^\Lambda_a \frac{dx^a}{dr}\bar{Z}(H)\Big) 
 &= & e^{K/2}\Big(\bar{Z}(H)\frac{dX^\Lambda}{dr}-Z(H)\frac{d\bar{X}^\Lambda}{dr}\Big) \cr
&+ &f^\Lambda (\bar{h}_\Sigma \partial_r H^\Sigma-\bar{f}^\Sigma \partial_r H_\Sigma)-\bar{f}^\Lambda (h_\Sigma \partial_r H^\Sigma-f^\Sigma \partial_r H_\Sigma)\ 
\label{eq:impeq3}
\end{eqnarray}
Using  the explicit expressions of the harmonic forms $H^\Lambda$ and $H_\Lambda$, we find 
\begin{eqnarray}
  2 i {\rm Im}\Big( f^\Lambda_a \frac{dx^a}{dr}\bar{Z}(H)\Big)  = 
  e^{K/2}\Big(\bar{Z}(H)\frac{dX^\Lambda}{dr}-Z(H)\frac{d\bar{X}^\Lambda}{dr}\Big)+\frac{1}{r^2}\big(\bar{Z}f^\Lambda-Z \bar{f}^\Lambda\big)\,
  \end{eqnarray}
Differentiating  \eqref{eq:ngst} with respect to $r$, we find 
\begin{equation}
   e^{K/2}\Big(\bar{Z}(H)\frac{dX^\Lambda}{dr}-Z(H)\frac{d\bar{X}^\Lambda}{dr}\Big) =   i \frac{p^\Lambda}{r^2}
   \end{equation}
and hence
\begin{eqnarray}
  2 i {\rm Im}\Big( f^\Lambda_a \frac{dx^a}{dr}\bar{Z}(H)\Big)  =   i \frac{p^\Lambda}{r^2}
+\frac{1}{r^2}\big(\bar{Z}f^\Lambda-Z \bar{f}^\Lambda\big)\, \label{eqlhs}
\end{eqnarray}

Let us now consider the radial variations of the moduli fields. Multiply both side of eq.(\ref{eq:nfo}) by $f^\Pi_a$ and using 
$g^{a\bar{b}}{f^\Lambda}_a{\bar{f}^\Sigma}_{\bar{b}} =-\frac{1}{2}(Im\mathcal{N})^{-1\Lambda\Sigma}-\bar{f}^\Lambda f^\Sigma$
and the central charge $Z=-2i(f^\Lambda Im\mathcal{N}_{\Lambda\Sigma}t^\Sigma)$ we find 
$$f^\Lambda_a \frac{dx^a}{dr}= \frac{e^U}{r^2}\frac{Z}{|Z|}(i t^\Lambda-Z\bar{f}^\Lambda) \ . $$ 
Multiplying both sides by $\bar{Z}(H)$, taking the imaginary part
\begin{equation}
2i Im \big(f^\Lambda_a \frac{dx^a}{dr}\bar{Z}(H)\big)=\frac{e^U}{r^2}\frac{Z\bar{Z}(H)}{|Z|}\big(i p^\Lambda-(Z \bar{f}^\Lambda-\bar{Z} f^\Lambda)\big)\label{eq:impeq1}
\end{equation}
 Using the metric ansatz \eqref{eq:mans} and reality condition of $Z(H)\bar{Z}$ we see the above relation agrees with \eqref{eqlhs}.

The stabilisation equations (\ref{eq:ngst}) are very similar to the attractor equations and can be obtained by suitably replacing 
the dyonic charges with appropriate harmonic functions. Thus, it is possible  to derive the exact solution for the flow equation
from the attractor solutions by substituting the harmonic functions appropriately. For the $D0-D4-D6$ system, we find
\begin{eqnarray}
e^{-2U} & = & \frac{2\chi_H}{H^0}\sqrt{1-\xi^2_H} 
\end{eqnarray}
with 
\begin{eqnarray}
x_1^a &=& \frac{1}{H^0}(H^a-\xi_H{I^a}_b H^b) \label{eq:sol1},\\
x_2^a &=& -\frac{\sqrt{1-\xi^2_H}}{H^0}{I^a}_b H^b\label{eq:sol2}
\end{eqnarray}
where $\xi_H=\frac{2D_H-H_0{H^0}^2}{2\chi_H}$ and $\chi_H={D_H}_a{I^a}_bH^b$. The expression for $D_H$ and 
$D_{H_a}$ are given by $D_{abc}H^aH^bH^c$ and $D_{abc}H^bH^c$ respectively.
For every non-trivial involution ${I^a}_b$ satisfying $D_{abc}{I^b}_e{I^c}_f = D_{aef}  $ we have
a new black hole solution. This relation has been solved for a two parameter in \cite{Manda:2015zoa} to obtain one
non-trivial involution in terms of the intersection numbers. Thus the most general $D0-D4-D6$ black hole in this case 
has two phases. It would be interesting to explore new black holes by solving the constraint on ${I^a}_b$ for more
general Calabi-Yau manifolds.  Moreover, one can generalise our analysis in the presence of all the dyonic charges 
and also for $N=2$ supergravity theories with more general pre-potential. For the non-supersymmetric attractors, 
the structure is richer and we have even multiple attractors with the same charge \cite{Dominic:2014zia}. It would be 
interesting to derive the analogues stabilisation and find the exact analytic expression for the corresponding black hole 
solutions in this case. 

Our analysis here to construct the black hole solutions has entirely based on the formalism developed in \cite{Ferrara:1997tw}.
Recently, new techniques were formulated by introducing new variables to rewrite the flow equations and black holes solutions 
were constructed by exploiting the symmetries of the equations of motion  \cite{Galli:2011fq,Meessen:2011aa,Galli:2012pt,Galli:2012ji,Bueno:2013pja}. In particular, they have been used to construct  black hole solutions in type $IIA$ compactification
in the presence of perturbative as well as non-perturbative corrections   \cite{Bueno:2012jc,Bueno:2013psa}.  It would be
interesting to understand the solutions constructed in the present work using the above formalism and study the effect of 
quantum corrections upon these black holes. We hope to explore some of these issues in future.


\begin{thebibliography}{99}
 
\bibitem{Strominger:1996sh} 
  A.~Strominger and C.~Vafa,
  Phys.\ Lett.\ B {\bf 379}, 99 (1996)
  doi:10.1016/0370-2693(96)00345-0
  [hep-th/9601029].
  
\bibitem{Cardy:1986ie} 
  J.~L.~Cardy,
  Nucl.\ Phys.\ B {\bf 270}, 186 (1986).
  doi:10.1016/0550-3213(86)90552-3

\bibitem{Ferrara:1995ih} 
  S.~Ferrara, R.~Kallosh and A.~Strominger,
  Phys.\ Rev.\ D {\bf 52}, 5412 (1995)
  doi:10.1103/PhysRevD.52.R5412
  [hep-th/9508072].
  
\bibitem{Strominger:1996kf} 
  A.~Strominger,
  Phys.\ Lett.\ B {\bf 383}, 39 (1996)
  doi:10.1016/0370-2693(96)00711-3
  [hep-th/9602111].

\bibitem{Moore:1998pn} 
  G.~W.~Moore,
  hep-th/9807087.
  
\bibitem{Moore:1998zu} 
  G.~W.~Moore,
  hep-th/9807056.

\bibitem{Manda:2015zoa} 
  T.~Mandal and P.~K.~Tripathy,
  Phys.\ Lett.\ B {\bf 749}, 221 (2015)
  doi:10.1016/j.physletb.2015.07.070
  [arXiv:1506.06276 [hep-th]].

\bibitem{Wijnholt:1999vk} 
  M.~Wijnholt and S.~Zhukov,
  hep-th/9912002.

\bibitem{Bellucci:2006xz} 
  S.~Bellucci, S.~Ferrara, M.~Gunaydin and A.~Marrani,
  Int.\ J.\ Mod.\ Phys.\ A {\bf 21}, 5043 (2006)
  doi:10.1142/S0217751X06034355
  [hep-th/0606209].
  
  
\bibitem{Tripathy:2005qp} 
  P.~K.~Tripathy and S.~P.~Trivedi,
  JHEP {\bf 0603}, 022 (2006)
  doi:10.1088/1126-6708/2006/03/022
  [hep-th/0511117].
  

\bibitem{Sabra:1997dh} 
  W.~A.~Sabra,
  Nucl.\ Phys.\ B {\bf 510}, 247 (1998)
  [hep-th/9704147].

\bibitem{Sabra:1997kq} 
  W.~A.~Sabra,
  Mod.\ Phys.\ Lett.\ A {\bf 12}, 2585 (1997)
  doi:10.1142/S0217732397002715
  [hep-th/9703101].
    
\bibitem{Mohaupt:2000mj} 
  T.~Mohaupt,
  Fortsch.\ Phys.\  {\bf 49}, 3 (2001)
  [hep-th/0007195].


  
\bibitem{Ferrara:1997tw} 
  S.~Ferrara, G.~W.~Gibbons and R.~Kallosh,
  Nucl.\ Phys.\ B {\bf 500}, 75 (1997)
  [hep-th/9702103].
  
\bibitem{Ferrara:2006yb} 
  S.~Ferrara, E.~G.~Gimon and R.~Kallosh,
  Phys.\ Rev.\ D {\bf 74}, 125018 (2006)
  [hep-th/0606211].
  
\bibitem{Ferrara:1996dd} 
  S.~Ferrara and R.~Kallosh,
  Phys.\ Rev.\ D {\bf 54}, 1514 (1996)
  doi:10.1103/PhysRevD.54.1514
  [hep-th/9602136].
  
  
\bibitem{Behrndt:1996jn} 
  K.~Behrndt, G.~Lopes Cardoso, B.~de Wit, R.~Kallosh, D.~Lust and T.~Mohaupt,
  Nucl.\ Phys.\ B {\bf 488}, 236 (1997)
  [hep-th/9610105].
  
\bibitem{Kallosh:2005ax} 
  R.~Kallosh,
  JHEP {\bf 0512}, 022 (2005)
  [hep-th/0510024].

\bibitem{Fre:1997jk} 
  P.~Fre,
  Nucl.\ Phys.\ Proc.\ Suppl.\  {\bf 57}, 52 (1997)
  doi:10.1016/S0920-5632(97)00353-8
  [hep-th/9701054].
  
\bibitem{Denef:2000nb} 
  F.~Denef,
  JHEP {\bf 0008}, 050 (2000)
  [hep-th/0005049].
  
\bibitem{Andrianopoli:2006ub} 
  L.~Andrianopoli, R.~D'Auria, S.~Ferrara and M.~Trigiante,
  Lect.\ Notes Phys.\  {\bf 737}, 661 (2008)
  [hep-th/0611345].
  

  
\bibitem{Dominic:2014zia} 
  P.~Dominic, T.~Mandal and P.~K.~Tripathy,
  JHEP {\bf 1412}, 158 (2014)
  [arXiv:1406.7147 [hep-th]].
 
  
  
  
\bibitem{Galli:2011fq} 
  P.~Galli, T.~Ortin, J.~Perz and C.~S.~Shahbazi,
  JHEP {\bf 1107}, 041 (2011)
  doi:10.1007/JHEP07(2011)041
  [arXiv:1105.3311 [hep-th]].
  
  
\bibitem{Meessen:2011aa} 
  P.~Meessen, T.~Ortin, J.~Perz and C.~S.~Shahbazi,
  Phys.\ Lett.\ B {\bf 709}, 260 (2012)
  doi:10.1016/j.physletb.2012.02.018
  [arXiv:1112.3332 [hep-th]].
  
\bibitem{Galli:2012pt} 
  P.~Galli, T.~Ortin, J.~Perz and C.~S.~Shahbazi,
  JHEP {\bf 1304}, 157 (2013)
  doi:10.1007/JHEP04(2013)157
  [arXiv:1212.0303 [hep-th]].
  
\bibitem{Galli:2012ji} 
  P.~Galli, P.~Meessen and T.~Ortin,
  JHEP {\bf 1305}, 011 (2013)
  doi:10.1007/JHEP05(2013)011
  [arXiv:1211.7296 [hep-th]].
  
  
\bibitem{Bueno:2013pja} 
  P.~Bueno, P.~Galli, P.~Meessen and T.~Ortin,
  JHEP {\bf 1309}, 010 (2013)
  doi:10.1007/JHEP09(2013)010
  [arXiv:1305.5488 [hep-th]].
  
  
\bibitem{Bueno:2012jc} 
  P.~Bueno, R.~Davies and C.~S.~Shahbazi,
  JHEP {\bf 1301}, 089 (2013)
  doi:10.1007/JHEP01(2013)089
  [arXiv:1210.2817 [hep-th]].
  
\bibitem{Bueno:2013psa} 
  P.~Bueno and C.~S.~Shahbazi,
  Class.\ Quant.\ Grav.\  {\bf 31}, 015023 (2014)
  doi:10.1088/0264-9381/31/1/015023
  [arXiv:1304.8079 [hep-th]].
 

  
\end{thebibliography}
\end{document}